\documentclass[shortnote]{jpsj3}

\usepackage{txfonts}
\usepackage{amsmath}
\usepackage{color}

\def\Pr247{Pr$_2$Ba$_4$Cu$_7$O$_{15-\delta}$}
\def\Tc{$T_\mathrm{c-onset}$}

\title{Nonmonotonic Pressure Dependence of the Lattice Parameter $a$ in the Quasi-one-dimensional Superconductor Pr$_2$Ba$_4$Cu$_7$O$_{15-\delta}$}

\author{Haruka Taniguchi$^1$\thanks{tanig@iwate-u.ac.jp}, Yuya Nakarokkaku$^1$, Riku Takahashi$^1$, Masatoshi Murakami$^1$, Atsuko Nakayama$^1$, Michiaki Matsukawa$^1$,
Satoshi Nakano$^2$, Makoto Hagiwara$^3$, and Takahiko Sasaki$^4$}
\inst{$^1$ Department of Physical Science and Materials Engineering, Iwate University, Morioka 020-8551, Japan \\
$^2$ National Institute for Materials Science (NIMS), Tsukuba, Ibaraki 305-0044, Japan \\
$^3$ Department of Comprehensive Sciences, Kyoto Institute of Technology, Kyoto 606-8585, Japan \\
$^4$ Institute for Materials Research, Tohoku University, Sendai 980-8577, Japan} 

\abst{
In the quasi-one-dimensional superconductor, Pr$_2$Ba$_4$Cu$_7$O$_{15-\delta}$, we found that the lattice parameter, $a$ increased against pressure, starting at 2.0~GPa at approximately 300~K.
This result suggests that the two-dimensional inter-conducting-chain interaction along the $a$ axis is most enhanced at approximately 2.0~GPa.
To discuss the dimensionality, we also measured the resistance under magnetic fields up to 14~T,
because magneto-resistance is induced when a system changes from one-dimensional to two-dimensional.
}

\begin{document}
\maketitle

\begin{figure}[htb]
\begin{center}
\includegraphics[width=3.3in]{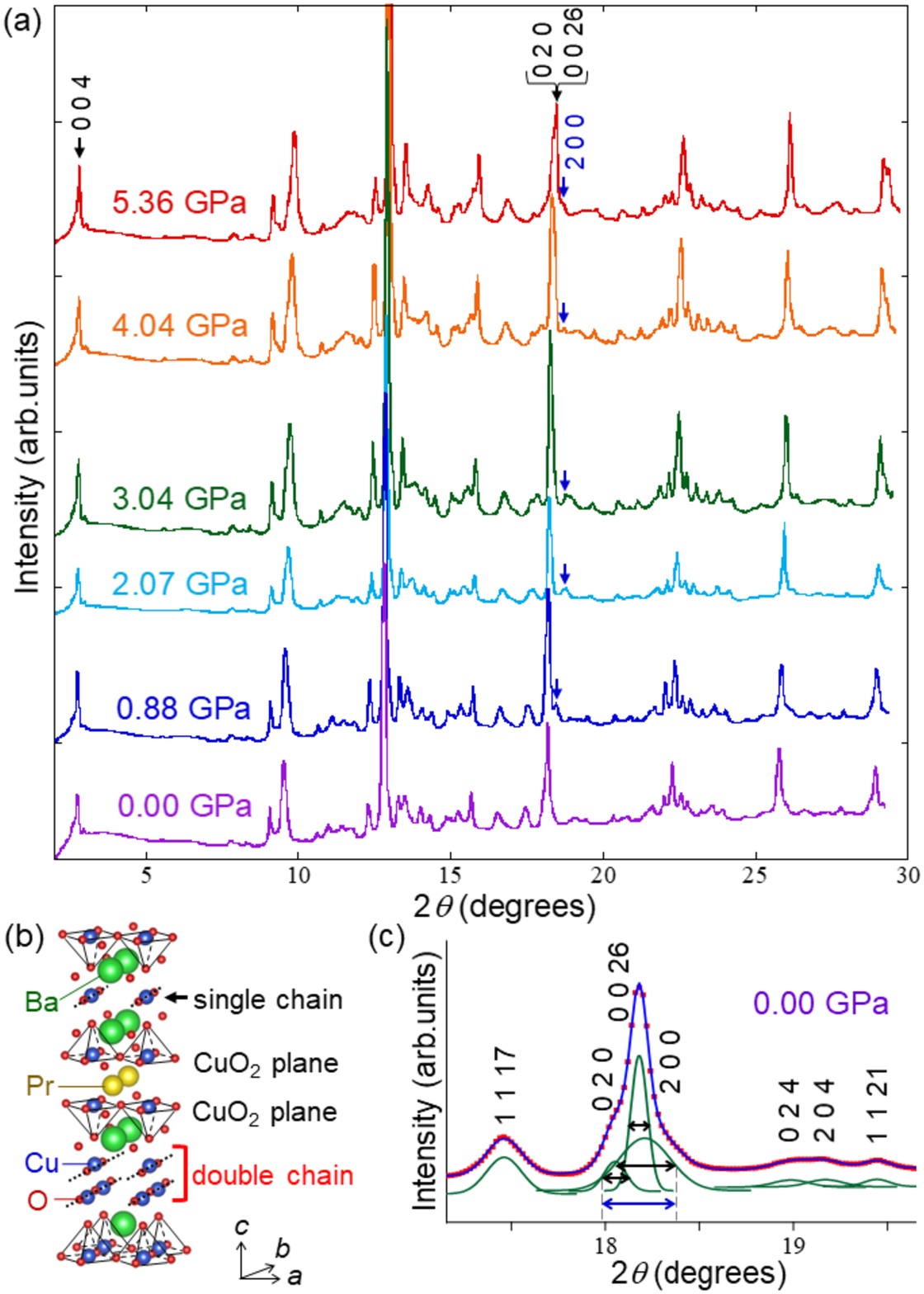}
\end{center}
\caption{(Color online)
(a) Superimposed display of powder XRD patterns of polycrystalline Pr$_2$Ba$_4$Cu$_7$O$_{15-\delta}$ ($\delta = 0.56$) observed at pressures of up to 5.36~GPa at approximately 300~K.
(b) Crystal structure of Pr$_2$Ba$_4$Cu$_7$O$_{15}$.
(c) Enlarged view of the main peaks (the 0 2 0, 0 0 26, and 2 0 0 reflection peaks) with fitted Gaussian functions.
}
\label{Fig1}
\end{figure}

Pr$_2$Ba$_4$Cu$_7$O$_{15-\delta}$ (Pr247)
(see crystal structure shown in Fig.~\ref{Fig1}(b))
exhibits superconductivity below approximately 27~K due to a reduction treatment in order to change the amount of oxygen deficiency~\cite{Matsukawa2004, Hagiwara2008}.
Electron conduction is realized along the Cu-O double chains~\cite{Sasaki2007}, which is different to Y-based cuprate superconductors with conductive CuO$_2$ planes.
The electron phase diagram of a quasi-one-dimensional (1D) system, such as Pr247 has been proposed in theoretical studies~\cite{Balents1996, Sano2005, Nishimoto2008}.
Regarding experiments, pressurization is expected to be effective in causing a dimensional crossover from 1D to 2D.
In our recent study on Pr247 ($\delta$ = 0.56) up to 1.6~GPa~\cite{Kuwabara2016}, pressure-induced magneto-resistance was found in the temperature region above the onset of superconductivity \Tc, 
suggesting the generation of quasi-2D warped Fermi surfaces.
The pressure dependence of the superconductivity was investigated for several samples with different $\delta$ values in magnetization or resistivity~\cite{Ishikawa2007, Ishikawa2009, Toshima2012, Kuwabara2016}.
In this study, to investigate changes in the dimensionality of Pr247 under pressure from the viewpoint of the lattice parameters, 
we measured x-ray diffraction under high pressure.
Furthermore, we report magneto-resistance below 2.0~GPa for the same sample as in Ref.~\citenum{Kuwabara2016}.

Polycrystalline Pr247 was synthesized via the citrate pyrolysis method~\cite{Hagiwara2006}.
The pelletized precursors were sintered at 887~$^\circ$C for 120~h under a 1-atm oxygen-gas atmosphere.
The reduction treatment was performed for the as-sintered samples at 500~$^\circ$C for 48 h in vacuum.
The value of the oxygen deficiency of the prepared sample was determined to be $\delta$ = 0.56 by gravimetric analysis.
The sample for XRD measurements was ground into a fine powder in an alumina mortar cooled in liquid nitrogen.
$In$-$situ$ XRD experiments under pressure were performed using a diamond anvil cell,
in which the powdered sample, ruby balls, and high-density He gas as pressure medium were set~\cite{Nakayama2014}.
Each pressure was determined by the ruby fluorescence method~\cite{Zha2000}.
Angle-dispersive powder patterns were obtained using synchrotron radiation (SR) at the beam line BL-18C at the Photon Factory, High Energy Accelerator Research Organization (KEK).
The SR beam was monochromatized to a wavelength of 0.6168~\AA~and introduced to the sample through a collimator with a diameter of 100~$\mu$m.
2D diffraction images were obtained by exposing the x-rays to the sample for 15~min using an imaging plate.
One image consists of 2000 $\times$ 2560 pixels with a resolution of 100 $\times$ 100 $\mu$m$^2$ per pixel.
The electrical resistance up to 14~T was measured in a piston-cylinder pressure cell
at the High Field Laboratory for Superconducting Materials (HFLSM), Tohoku University~\cite{Kuwabara2016}, whereby Fluorinert was used as the pressure medium.
The magnitude of pressure was calibrated using the superconducting transition temperature of lead.

Figure~\ref{Fig1}(a) shows the SR XRD patterns of Pr247 with an oxygen deficiency determined to be $\delta = 0.56$.
The pattern profiles showed an orthorhombic $Ammm$ symmetry.
As shown in Fig.~\ref{Fig1}(c), the lattice parameters $a$, $b$, and $c$ were obtained by fitting Gaussian functions to the 2 0 0, 0 2 0, and 0 0 26 reflection peaks with the least square method.
In general, except for some cases mentioned below,
the magnitude of each error was estimated from the full width at half maximum (FWHM) of each fitted Gaussian function 
(see black arrows in Fig.~\ref{Fig1}(c)).
We confirmed that the parameter $c$ determined from the 0 0 26 reflection exhibited the same pressure dependence as that obtained from the 0 0 4 reflection.
The accuracy of the obtained value of $a$ is enhanced above 0.88~GPa, 
because the 2 0 0 peak was isolated from the 0 2 0 and 0 0 26 peaks, as indicated by the blue arrows in Fig.~\ref{Fig1}(a).
We evaluated the error in $b$ carefully, 
because the low 0 2 0 peak overlapped with the tail of the high 0 0 26 peak and the top position of the 0 2 0 peak was difficult to be distinguished clearly:
As shown by the blue arrow in Fig.~\ref{Fig1}(c),
we defined the error of $2\theta$ for the 0 2 0 reflection to cover not only the FWHM region of the 0 2 0 Gaussian function but also that of the 0 0 26 Gaussian function
(and also, that of the 2 0 0 Gaussian function for $P <$ 0.88~GPa).
For $P <$ 0.88~GPa, the error of $a$ was estimated in the same manner.

\begin{figure}[htb]
\begin{center}
\includegraphics[width=3.2in]{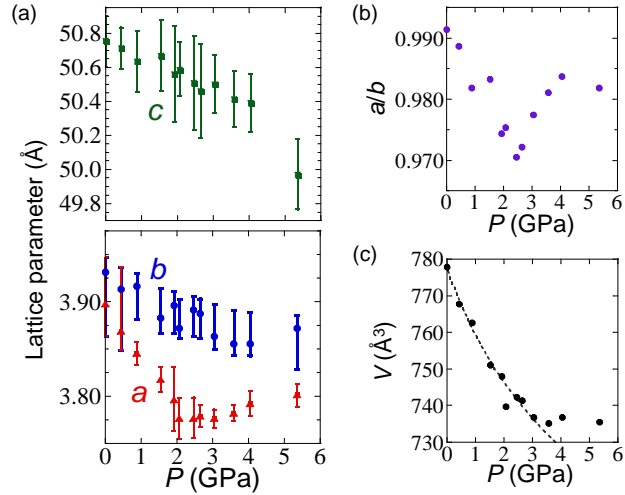}
\end{center}
\caption{(Color online)
Pressure dependence of (a) the lattice parameters $a$, $b$, and $c$, (b) the ratio of $a$/$b$, and (c) the unit cell volume of Pr$_2$Ba$_4$Cu$_7$O$_{15-\delta}$ ($\delta = 0.56$).
The broken line in panel (c) represents a Murnaghan-type equation of state fit to the experimental data up to 3~GPa~\cite{Murnaghan1944}.
}
\label{Fig2}
\end{figure}

Figure~\ref{Fig2}(a) presents the pressure dependence of the lattice parameters.
We obtained $a = 3.90(4)$~\AA, $b = 3.932(16)$~\AA, and $c = 50.75(15)$~\AA~at ambient pressure,
and $a = 3.801(12)$~\AA, $b = 3.872(14)$~\AA, and $c = 50.0(2)$~\AA~ at 5.36~GPa.
Interestingly, the length of $a$ started to increase at pressures above 2.0~GPa.
As shown in Fig.~\ref{Fig2}(b), the ratio of $a/b$ exhibited a minimum at 2.5~GPa.
Because Cu-O double chains exist along the $b$ axis and the length of $a$ is the distance between neighboring double chains (Fig.~\ref{Fig1}(b)),
the suppression of $a$ indicates an enhancement in inter-double-chain interactions along the $a$ axis, which is related to quasi-2D conductivity.
Thus, the nonmonotonic pressure dependence of $a$ suggests that the quasi-2D feature is most enhanced at approximately 2.0~GPa.
As shown in Fig.~\ref{Fig2}(c), the pressure dependence of the unit cell volume below 3.0~GPa could be described well by a Murnaghan-type equation of state~\cite{Murnaghan1944}
$ P = (B_0 / B'_0) ((L / L_0)^{-B'_0} - 1) $,
and the bulk modulus $B_0$ was determined to be 33.9~GPa, which is similar to that of cuprates~\cite{Ledbetter1990, Hyatt2001}.
Although a structural transition was not observed, deviations from the Murnaghan-type curve above 3.0~GPa suggest 
that the lattice deformation at high pressures is different from that at low pressures.

\begin{figure}[htb]
\begin{center}
\includegraphics[width=3.2in]{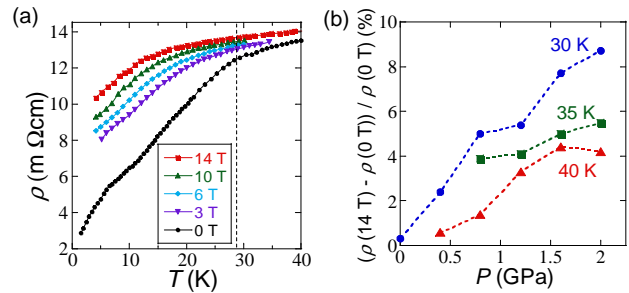}
\end{center}
\caption{(Color online)
(a)Resistivity of Pr$_2$Ba$_4$Cu$_7$O$_{15-\delta}$ ($\delta = 0.56$) under several magnetic fields at 2.0~GPa. 
The broken line indicates \Tc.
(b) Pressure dependence of the magneto-resistance at 14~T. For $P \leq$ 1.6~GPa, our previous paper~\cite{Kuwabara2016} was referred to.
}
\label{Fig3}
\end{figure}

As shown in Fig.~\ref{Fig3}(a), we observed magneto-resistance above \Tc~at 2.0~GPa.
Magneto-resistance is important for discussing dimensionality.
1D systems do not exhibit magneto-resistance because the Lorentz force cannot bend the orbitals of the itinerant electrons owing to the straight Fermi surfaces without the wave numbers being perpendicular to the conducting direction,
whereas 2D systems exhibit magneto-resistance because of warped Fermi surfaces with finite wave numbers perpendicular to the conducting direction.
As shown in Fig.~\ref{Fig3}(b), the magneto-resistance at 30 and 35~K was enhanced with pressure up to 2.0~GPa, and that at 40~K exhibited a maximum at approximately 1.6~GPa.
Although we were not able to apply a higher pressure owing to the limits of the piston-cylinder cell, 
the magneto-resistance above 2.0~GPa is an intriguing future issue for investigation 
because the nonmonotonic pressure dependence of $a$ predicts that the magneto-resistance would be suppressed above approximately 2.0~GPa owing to weakened inter-double-chain interactions along the $a$ axis.

In summary, we performed synchrotron x-ray diffraction and magneto-resistance measurements of the metallic double-chain superconductor Pr247 
in the pressure range of 0.88-5.36~GPa and 0.4-2.0~GPa, respectively.
Interestingly, the lattice parameter $a$ was elongated against pressure above 2.0~GPa.
This result suggests that the pressure-induced magneto-resistance observed below 2.0~GPa would be suppressed at higher pressures.

\begin{acknowledgment}
This work was supported by Iwate University, Photon Factory (Project 2017P005), NIMS, and HFLSM.
\end{acknowledgment}


\bibliography{string,Pr247,Others}

\end{document}